\documentclass[sigconf,natbib=true,anonymous=false]{acmart}
\renewcommand\footnotetextcopyrightpermission[1]{} 

\usepackage{enumitem}
\usepackage{epstopdf}
\usepackage{multirow}
\usepackage{balance}
\usepackage[ruled,linesnumbered]{algorithm2e}
\usepackage{caption}
\usepackage{subcaption}
\usepackage{graphicx}
\usepackage{float}
\usepackage{adjustbox}
\usepackage{wrapfig,lipsum}
\usepackage{booktabs,etoolbox}
\usepackage{amsmath,amsfonts,amsthm,bm}
\usepackage{layouts}
\usepackage{cleveref}
\usepackage{tablefootnote}
\usepackage{makecell}
\usepackage{upgreek}
\usepackage{mathtools, stmaryrd}
\usepackage{algpseudocode}
\usepackage{xparse}

\AtBeginDocument{%
  \providecommand\BibTeX{{%
    \normalfont B\kern-0.5em{\scshape i\kern-0.25em b}\kern-0.8em\TeX}}}

\setcopyright{acmcopyright}
\copyrightyear{2018}
\acmYear{2018}
\acmDOI{XXXXXXX.XXXXXXX}

\acmConference[Conference acronym 'XX]{Make sure to enter the correct
  conference title from your rights confirmation emai}{June 03--05,
  2018}{Woodstock, NY}
\acmBooktitle{Woodstock '18: ACM Symposium on Neural Gaze Detection,
 June 03--05, 2018, Woodstock, NY} 
\acmPrice{15.00}
\acmISBN{978-1-4503-XXXX-X/18/06}

\begin{document}

\title{Cross-lingual Knowledge Transfer via Distillation for Multilingual Information Retrieval}

\author{Zhiqi Huang}
\affiliation{%
  \institution{University of Massachusetts Amherst}
  \city{Amherst}
  \state{MA}
  \country{USA}
}
\email{zhiqihuang@cs.umass.edu}

\author{Puxuan Yu}
\affiliation{%
  \institution{University of Massachusetts Amherst}
  \city{Amherst}
  \state{MA}
  \country{USA}
}
\email{pxyu@cs.umass.edu}

\author{James Allan}
\affiliation{%
  \institution{University of Massachusetts Amherst}
  \city{Amherst}
  \state{MA}
  \country{USA}
}
\email{allan@cs.umass.edu}

\renewcommand{\shortauthors}{Huang et al.}

\begin{abstract}
In this paper, we introduce the approach behind our submission for the MIRACL challenge, a WSDM 2023 Cup competition that centers on ad-hoc retrieval across 18 diverse languages. Our solution contains two neural-based models. The first model is a bi-encoder re-ranker, on which we apply a cross-lingual distillation technique to transfer ranking knowledge from English to the target language space. The second model is a cross-encoder re-ranker trained on multilingual retrieval data generated using neural machine translation. We further fine-tune both models using MIRACL training data and ensemble multiple rank lists to obtain the final result. According to the MIRACL leaderboard, our approach ranks 8th for the Test-A set and 2nd for the Test-B set among the 16 known languages.
\end{abstract}



\keywords{WSDM Cup; multilingual retrieval; knowledge distillation}

\maketitle
\section{Introduction}  \label{sec:introduction}
MIRACL (Multilingual Information Retrieval Across a Continuum of Languages)~\cite{zhang2022making} is a multilingual dataset that includes 18 languages and is intended for ad hoc retrieval tasks. It is also one of the challenges in WSDM 2023 Cup\footnote{\url{https://project-miracl.github.io/}}. The primary objective of MIRACL is to evaluate the effectiveness of monolingual retrieval systems in diverse linguistic environments. There are two tracks in the dataset. The known-languages track covers 16 languages, and participants are provided with a list of language names and MIRACL training data, including monolingual queries and human-annotated documents for each language. In the surprise-languages track, there are two undisclosed languages, German and Yoruba, revealed only in the last week of the submission deadline. No training data are provided for this track. A small set of development data is available for each language to fine-tune system parameters. The nDCG@10 metric is employed to evaluate retrieval performance in MIRACL. 

Along with the data, MIRACL also provides three baselines that serve as a reference for participants to compare and build upon in their approaches. The first baseline is BM25, a lexical matching algorithm. The second baseline is the multilingual dense passage retriever (mDPR)~\cite{zhang2021mr}, which is a neural-based bi-encoder retrieval model focused on semantic matching. The third is a combination of the first and second, named Hybrid. Since, on average, Hybrid achieved 88.9\% recall in the top-100 retrieved documents, we employ a two-stage retrieval approach to tackle MIRACL, where first we adopt the rank lists produced by Hybrid for each language as the initial rankings and then re-rank the initial set of candidate documents using neural re-rankers.
In the re-ranking stage, our solution contains two neural-based models. The first model is a bi-encoder re-ranker, on which we apply a cross-lingual distillation technique~\cite{huang2023improving} to transfer ranking knowledge from English to the target language. More specifically, using bitext, we form the token alignment task as an optimal transportation problem to distill the knowledge from a well-trained English model into the target language. 
The second model is a cross-encoder re-ranker trained on multilingual retrieval data generated with neural machine translation. 
We further fine-tune both re-ranking models using MIRACL training data and ensemble multiple rank lists from both retrieval stages to obtain the final result.

\section{Background}  \label{sec:background}
\begin{figure*}[t]
    \captionsetup[subfigure]{font=footnotesize,labelfont=footnotesize}
    \begin{subfigure}[t]{0.47\textwidth}
        \raggedleft
        \includegraphics[width=\linewidth]{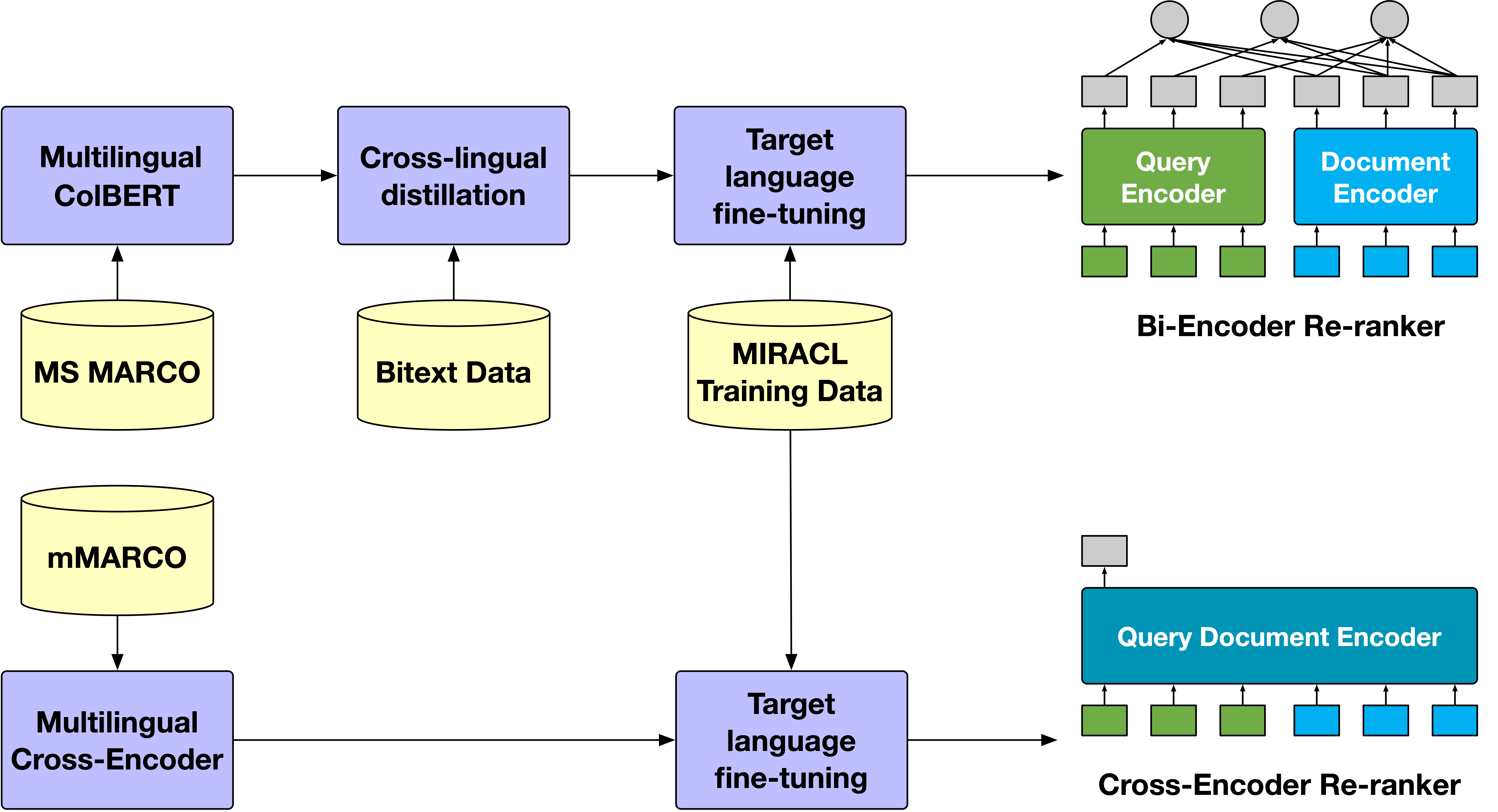}
        \caption{Model training pipeline.}
        \label{fig:train-flow}
    \end{subfigure}
    \hfill
    \begin{subfigure}[t]{0.47\textwidth}
        \raggedright
        \includegraphics[width=\linewidth]{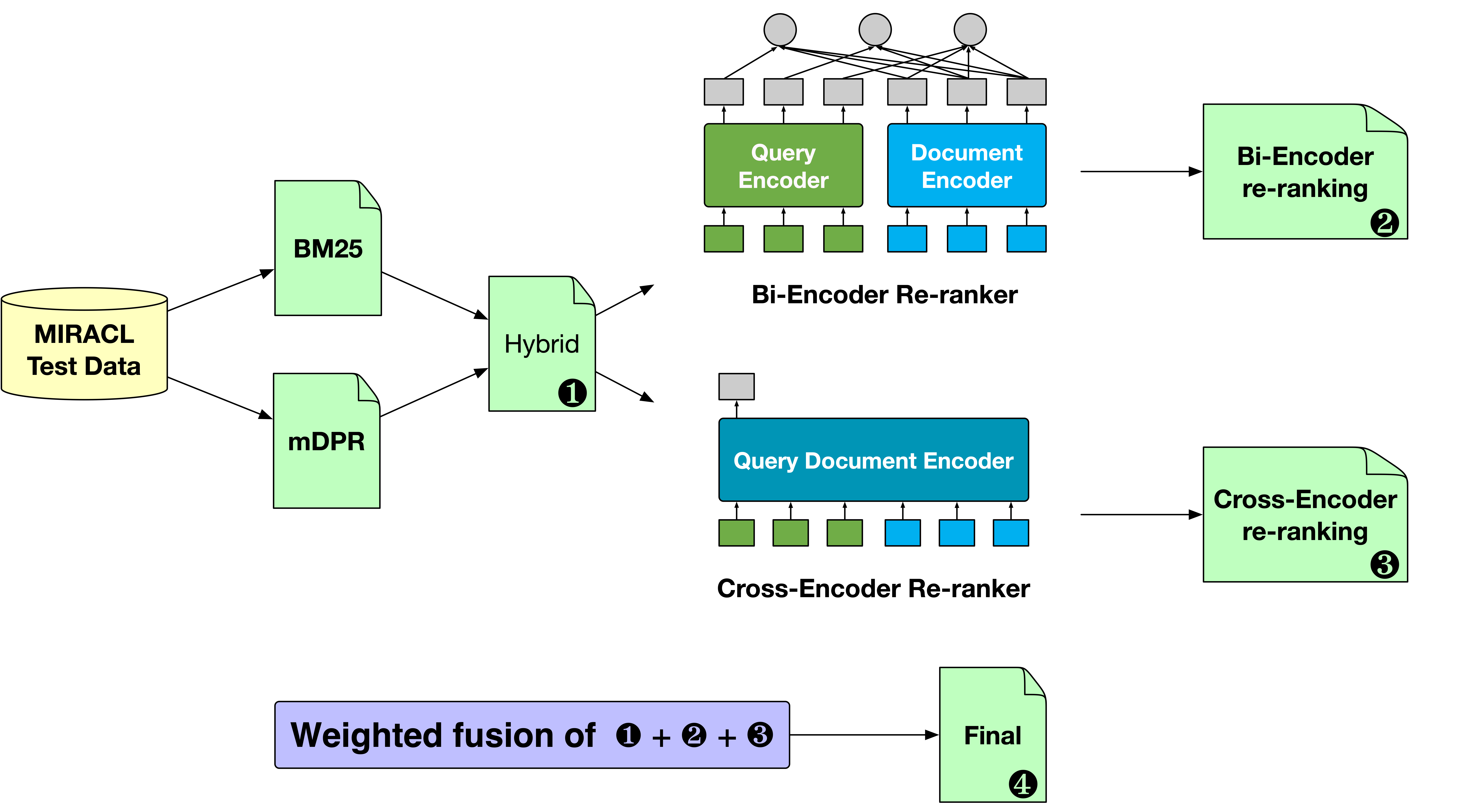}
        \caption{Model evaluation pipeline.}
        \label{fig:test-flow}
    \end{subfigure}
\caption{Workflows of Our Proposed Solution}
\label{fig:work-flow}
\end{figure*}

\subsection{Neural Multilingual Models for Retrieval}
Previous neural models for retrieval involving multiple languages rely on the combination of multilingual word embeddings and neural matching models~\cite{yu2020study}.
Benefiting from transformer-based pre-trained language models, neural ranking models have made significant progress. The advent of multilingual pre-trained language models~\cite{conneau-etal-2020-unsupervised} provides the possibility of jointly learning downstream applications for many languages with the same model.
For cross-encoder retrieval architecture~\cite{huang2021mixed,yu2021cross}, the model takes the concatenation of the query and document as input. An embedding produced for the prepended special ``[CLS]'' token is fed into a feed-forward layer to produce a score for the input pair~\cite{nogueira2019passage}. Since each query-document pair needs encoding during inference, retrieval based on cross-encoders tends to be computationally demanding. As a result, such models typically rely on sparse retrieval methods based on lexical matching as the initial step for retrieving relevant information. 
Dense retrieval based on bi-encoder architecture is proposed to overcome the sparse retrieval bottleneck~\cite{luan2021sparse, lee2019latent, karpukhin2020dense}. With the separation of query and document encoders, dense retrieval has already shown success on many retrieval tasks~\cite{khattab2020colbert,gao2021unsupervised}, including multilingual retrieval~\cite{zhang2021mr}. In our solution, we leverage MIRACL baselines as the initial ranking step and apply both cross-encoder and bi-encoder for re-ranking. 

\subsection{Knowledge Distillation}
Proposed by \citet{hinton2015distilling}, knowledge distillation is a method to train a model, called the student, using valuable information provided by the output of another model, called the teacher. This way, the teacher model's knowledge can be transferred into the student model. The idea of knowledge distillation is wildly used in the field of computer vision~\cite{xie2020self, yuan2019ckd, Lin_2022_CVPR}, natural language processing~\cite{sanh2019distilbert, reimers2020making} and information retrieval~\cite{lu2020twinbert, hofstatter2020improving, li-etal-2022-learning-cross}.
A typical framework for knowledge distillation relies on a teacher model to directly generate a target distribution~\cite{gou2021knowledge,ma2022deep}. 
Our solution adopts the OPTICAL~\cite{huang2023improving} framework to perform cross-lingual knowledge distillation, which casts the cross-lingual token alignment task as an optimal transport problem to transfer retrieval knowledge from English (which has the most resources for relevance) to other languages.

\section{First Stage Retrieval}  \label{sec:first-stage}

\begin{table}[t]
    \centering
    \captionsetup{width=\linewidth}
    \caption{Average performance of MIRACL baselines.}
    \label{tab:baseline}
    \begin{adjustbox}{width=0.45\textwidth}
    \renewcommand{\arraystretch}{1.2}
    \begin{tabular}{lcccccc}
        \toprule
        \multirow{2}{*}{\textbf{Baseline}}& \multicolumn{3}{c}{\textbf{nDCG@10}} & \multicolumn{3}{c}{\textbf{Recall@100}} \\
        & BM25 & mDPR & Hybrid & BM25 & mDPR & Hybrid \\
        \cmidrule(lr){1-2} \cmidrule(lr){2-4} \cmidrule(lr){5-7}
        Average & 0.393 & 0.415 & 0.578 & 0.787 & 0.788 & 0.889 \\
        \bottomrule
    \end{tabular}
    \end{adjustbox}
\end{table}

To address the MIRACL challenge, we have employed a two-stage retrieval approach. In the first stage, we utilize the rank lists provided as the MIRACL baselines. Table 1 shows the average performance of the three baselines provided by the MIRACL project. We can see that under zero-shot and out-of-distribution (OOD) settings, BM25 still demonstrates a strong performance compared to neural methods. Furthermore, the combination of lexical and semantic matching methods (Hybrid) can improve both precision and recall. Since Hybrid achieves an average recall of 88.9\% for the top 100 documents, we use it as our initial retrieval step and focus on the re-ranking step.

\section{Methodology}  \label{sec:methodology}

\begin{table*}[t]
\centering
\captionsetup{width=0.98\linewidth}
\caption{Results on development split of 16 known languages. The neural models re-rank the top-100 results produced by Hybrid. The highest value is marked with bold text.}
\label{tab:dev-results}
\begin{adjustbox}{width=0.95\textwidth}
\renewcommand{\arraystretch}{1.2}
\begin{tabular}{lccccccccccccccccc}
\toprule
 \multirow{2}{*}[-2pt]{\textbf{\shortstack[l]{Retrieval\\Methods}}} & \multicolumn{16}{c}{\textbf{nDCG@10}} \\
 \cmidrule(lr){2-18}
 &  Avg. &  ar & bn & en & es & fa & fi & fr & hi & id & ja & ko & ru & sw & th & te & zh  \\
 \midrule
BM25 & 0.393 & 0.481 & 0.508 & 0.351 & 0.319 & 0.333 & 0.551 & 0.183 & 0.458 & 0.449 & 0.369 & 0.419 & 0.334 & 0.383 & 0.494 & 0.484 & 0.180 \\
mDPR & 0.415 & 0.499 & 0.443 & 0.394 & 0.478 & 0.48 & 0.472 & 0.435 & 0.383 & 0.272 & 0.439 & 0.419 & 0.407 & 0.299 & 0.356 & 0.358 & 0.512 \\
Hybrid & 0.578 & 0.673 & 0.654 & 0.549 & 0.641 & 0.594 & 0.672 & 0.523 & 0.616 & 0.443 & 0.576 & 0.609 & 0.532 & 0.446 & 0.602 & 0.599 & 0.526 \\
 \midrule
 mColBERT & 0.671 & 0.776 &  0.768 &  0.566 &  0.590 &  0.595 &  0.766 &  0.576 &  0.632 &  0.589 &  0.678 &  0.698 &  0.646 &  0.674 &  0.795 &  0.764 &  0.628 \\
 CrossEncoder & 0.732 & 0.820 &  0.817 &  0.665 &  0.681 &  0.659 &  0.804 &  0.637 &  0.702 &  0.641 &  0.749 &  0.786 &  0.719 &  0.694 &  0.810 &  0.807 &  0.721 \\
 Fusion & \textbf{0.764} &  \textbf{0.842} &  \textbf{0.837} &  \textbf{0.689} &  \textbf{0.761} &  \textbf{0.714} &  \textbf{0.823} &  \textbf{0.706} &  \textbf{0.763} &  \textbf{0.654} &  \textbf{0.763} &  \textbf{0.801} &  \textbf{0.741} &  \textbf{0.705} &  \textbf{0.818} &  \textbf{0.823} &  \textbf{0.791} \\

\bottomrule
\end{tabular}
\end{adjustbox}
\end{table*}

In this section, we introduce the model training pipeline of our solution which contains two neural models built in multiple steps. Figure~\ref{fig:train-flow} depicts the workflow for creating two neural re-rankers. For the bi-encoder model, we follow the ColBERT~\cite{khattab2020colbert} architecture and initialize model parameters using a multilingual pre-trained language model. We first tune the model on MS MARCO~\cite{nguyen2016ms} passage ranking data. Then we apply cross-lingual knowledge distillation~\cite{huang2023improving} with bitext data to transfer knowledge from English to each target language. For the cross-encoder architecture, we adopt a model checkpoint trained on mMARCO~\cite{bonifacio2021mmarco} dataset. Finally, we further fine-tune both models on MIRACL training data. Following, we describe each step in detail.

\subsection{Model Training}
\noindent \textbf{Bi-encoder Model}. Following the ColBERT~\cite{khattab2020colbert} model architecture, we first build a multilingual ColBERT (mColBERT) $M$ contains two components -- query encoder $E_{M_q}$ and document encoder $E_{M_d}$. Given a query $q$ and a candidate document $d$, the matching score between $q$ and $d$, denoted $S_{q,d}$, is computed as:
\begin{align}
\label{eq:score}
   S_{q,d} = \sum\limits_{i=1}^{|q|} \max\limits_{j=1}^{|d|} \ E_{M_q}(q_i) \cdot E^T_{M_d}(d_j)
\end{align}
where $E_{M_q}(q_i)$ is the $i$-th token representation of the query and $E_{M_d}(d_j)$ is the $j$-th token representation of the document.
The scoring function applies the \textit{maxsim} operation on each query token to softly search against all document tokens to find the best token that reflects its context and then sums over all the query tokens. Despite the separation of the query and document encoding process, ColBERT is a Siamese neural network in which query and document encoders share the same weights. We initialize the encoder parameters using XLM-R, a multilingual pre-trained language model, and tune the model for the retrieval task on MS MARCO passage ranking dataset. More specifically, given a triplet of \{query, relevant passage, non-relevant passage\}, the models are trained using pairwise cross-entropy loss. Although the knowledge learned for retrieval is based on English data only, with the help of multilingual initialization, the model is capable of conducting retrieval tasks in different languages in the zero-shot setting.

\noindent \textbf{Cross-encoder Model.} \citet{bonifacio2021mmarco} built a multilingual passage ranking dataset, mMARCO, by translating the queries and passages in MS MARCO into the target language using neural machine translation (NMT). Since mMARCO provides retrieval data in 13 languages and 9 of which overlap with MIRACL languages, we build our cross-encoder (CrossEncoder) using triples in mMARCO. Instead of training our model checkpoint, we direct adopt a shared cross-encoder checkpoint from Huggingface\footnote{\url{https://huggingface.co/cross-encoder/mmarco-mMiniLMv2-L12-H384-v1}} contributed by SBERT\footnote{\url{https://www.sbert.net/}}.

\subsection{Cross-lingual Distillation}
To improve the performance of the bi-encoder model on different target languages, we employ the OPTICAL~\cite{huang2023improving} framework, a cross-lingual knowledge distillation method, to transfer retrieval knowledge learned from English to each of the MIRACL languages. 
The intuition is to align the target language token embeddings with the English embeddings. Since the English embeddings contain the knowledge for retrieval, the retrieval performance on the target language can be enhanced once their embeddings are successfully aligned. 
The framework builds a new student model focusing on a specific language by distilling knowledge from the teacher model, a well-trained English retrieval model.
This technique only requires bitext data for training, thus applicable to a wide range of languages, especially low-resource languages.
For each language, we sample up to 2M bitext data from CCAligned~\cite{el-kishky-etal-2020-ccaligned} dataset and apply OPTICAL framework to build a student model using $M$ as the teacher. 

\subsection{Fine-tuning and Evaluation}
We further fine-tune both bi-encoder and cross-encodes models on MIRACL training data. To best train our model under the re-ranking setting, we sample both relevant and non-relevant examples from the top 100 results retrieved by Hybrid. We apply the Localized Contrastive Estimation (LCE)~\cite{pradeep2022squeezing} as the loss function to incorporate in-distribution hard negatives. More specifically, instead of sampling one negative document per query, we sample 6 negative examples from the top-ranked documents in the first-stage result (Hybrid). The loss function scores the positive instance and multiple negative instances and encourages the model to score the positive higher than all the negatives.

Once fine-tuned, our approach has a bi-encoder and cross-encoder models for each language. Given the test queries, we first follow the same baseline procedure to generate the Hybrid rank list as the first-stage retrieval, and then re-rank the top 100 documents with two neural re-rankers, respectively. Finally, we combine three rank lists (Hybrid, bi-encoder, and cross-encoder) into one final rank list (Fusion) with a weighted sum of the ranking scores. The weight for each rank list is tuned based on the development set. Note that the scores from different methods are first min-max normalized to [0, 1]. The complete evaluation pipeline is shown in Figure~\ref{fig:test-flow}.

\section{Results and Discussion}  \label{sec:results}

\begin{table}[t]
    \centering
    \captionsetup{width=\linewidth}
    \caption{Average performance on two test splits.}
    \label{tab:test-results}
    \begin{adjustbox}{width=0.35\textwidth}
    \renewcommand{\arraystretch}{1.2}
    \begin{tabular}{lcccc}
        \toprule
        \multirow{2}{*}[-2pt]{\shortstack[l]{\textbf{Test Splits} \\ (Private Score)}} & \multicolumn{4}{c}{\textbf{nDCG@10}} \\
        & BM25 & mDPR & Hybrid & Fusion \\
        \midrule
        Test-A & 0.449 & 0.398 & 0.635 & \textbf{0.759} \\
        Test-B & 0.347 & 0.378 & -- & \textbf{0.718} \\
        \bottomrule
    \end{tabular}
    \end{adjustbox}
\end{table}

Our experimental results are presented in Table 2 and Table 3. In Table 2, we report the performance of each compared method split by languages on the development split. We can see that both re-rankers outperform the first-stage retrieval method. Comparing mColBERT with CrossEncoder, we can see that cross-encoder architecture with all-to-all interaction between query and document has the advantage over bi-encoder where the interaction between query and document representations is delayed to the scoring phase for efficiency. Finally, the weighted sum fusion of the two re-ranking lists with the Hybrid initial rank list gives the best performance. The Fusion results indicate that each method captures distinct aspects of query-document matching, and collectively they all contribute to the final performance. 
Moreover, we can see a consistent performance between development and Test-A splits, yet a significant drop from Test-A to Test-B. We suspect the performance gap is due to the distribution shift, which needs further investigation.

\section{Conclusion}  \label{sec:conclusion}
This paper presents a comprehensive solution to the Multilingual Information Retrieval Across a Continuum of Languages (MIRACL) challenge of the WSDM CUP 2023. We leverage the baseline provided by MIRACL project as the first-stage retrieval method and focus on the re-ranking step. Our model-building pipeline comprises three components: First, we pre-fine-tune a multilingual language model on retrieval data to build both bi-encoder and cross-encoder retrieval models. Then for the bi-encoder, we apply a cross-lingual knowledge distillation framework to transfer knowledge learned by English data to the target language. Lastly, we fine-tune both re-rankers using MIRACL training data with localized contrastive estimation. At test time, the best performance is a fusion of results from neural re-rankers with the first-stage rank list. In future work, we are interested in exploring the performance of our bi-encoder model under an end-to-end setting.

\begin{acks}
We express our sincere appreciation to the MIRACL competition organizers for their hard work and valuable contributions throughout the event. We are grateful for their efforts in hosting a superb competition and providing top-notch datasets. We also want to thank all those involved in sponsoring the WSDM 2023 for their support and commitment, which have contributed to the success of this competition.
\end{acks}

\bibliographystyle{ACM-Reference-Format}
\bibliography{01-reference}
\end{document}